\documentclass[12pt,preprint]{aastex}
\usepackage{color}

\slugcomment{To appear in ApJS.} \shorttitle{The White Mountain
Polarimeter} \shortauthors{Levy et al.}

\begin{document}

\title{The White Mountain Polarimeter Telescope and an Upper Limit on CMB Polarization}

\author{Alan R. Levy,\altaffilmark{1,2,3,4} Rodrigo Leonardi,\altaffilmark{1,5}
Markus Ansmann,\altaffilmark{1} Marco Bersanelli,\altaffilmark{6}
Jeffery Childers,\altaffilmark{1,3} Terrence D.
Cole,\altaffilmark{1} Ocleto D'Arcangelo,\altaffilmark{7} G. Vietor
Davis,\altaffilmark{1} Philip M. Lubin,\altaffilmark{1,3} Joshua
Marvil,\altaffilmark{1,3,8} Peter R. Meinhold,\altaffilmark{1,2,3}
Gerald Miller,\altaffilmark{1,9} Hugh O`Neill,\altaffilmark{1,3}
Fabrizio Stavola,\altaffilmark{10} Nathan C.
Stebor,\altaffilmark{1,2,3} Peter T. Timbie,\altaffilmark{11}
Maarten van der Heide,\altaffilmark{1} Fabrizio
Villa,\altaffilmark{10} Thyrso Villela,\altaffilmark{5} Brian D.
Williams,\altaffilmark{1,2,3} and Carlos A.
Wuensche\altaffilmark{5}}

\email{Alan\_Levy@Raytheon.com}

\altaffiltext{1}{Department of Physics, University of California,
Santa Barbara, CA 93106} \altaffiltext{2}{University of California,
White Mountain Research Station, Bishop, CA 93514}
\altaffiltext{3}{University of California, Santa Barbara Center for
High Altitude Astrophysics at White Mountain, University of
California, Santa Barbara, CA 93106} \altaffiltext{4}{Now at
Raytheon Vision Systems, Goleta, CA 93117}
\altaffiltext{5}{Instituto Nacional de Pesquisas Espaciais,
Divis\~{a}o de Astrof\'{i}sica, Caixa Postal 515, S\~{a}o Jos\'{e}
dos Campos, SP 12210-070, Brazil} \altaffiltext{6}{Dipartimento di
Fisica, Universit\`{a} degli Studi di Milano, Via Celoria, 16, 20133
Milano, Italy} \altaffiltext{7}{IFP-CNR, Via Roberto Cozzi, 53,
20125 Milano, Italy} \altaffiltext{8}{Now at Department of Physics,
New Mexico Institute of Mining and Technology, Socorro, NM 87801}
\altaffiltext{9}{Now at California Institute of Technology,
Pasadena, CA 91125}\altaffiltext{10}{INAF-IASF Bologna, Via P.
Gobetti, 101, 40129 Bologna, Italy}\altaffiltext{11}{Department of
Physics, University of Wisconsin, Madison, WI 53706}

\begin{abstract}
The White Mountain Polarimeter (WMPol) is a dedicated ground-based
microwave telescope and receiver system for observing polarization
of the Cosmic Microwave Background. WMPol is located at an altitude
of 3880 meters on a plateau in the White Mountains of Eastern
California, USA, at the Barcroft Facility of the University of
California White Mountain Research Station. Presented here is a
description of the instrument and the data collected during April
through October 2004. We set an upper limit on $E$-mode polarization
of 14 $\mu\mathrm{K}$ ($95\%$ confidence limit) in the multipole
range $170<\ell<240$. This result was obtained with 422 hours of
observations of a 3 $\mathrm{deg}^2$ sky area about the North
Celestial Pole, using a 42 GHz  polarimeter. This
upper limit is consistent with $EE$ polarization predicted from a
standard $\Lambda$-CDM concordance model.
\end{abstract}


\section{Introduction}

Polarization of the Cosmic Microwave Background (CMB) results from
Thomson scattering of CMB photons from local quadrupolar temperature
anisotropies during last scattering. Measurements of CMB
polarization promise to help to constrain cosmological models, as
discussed in, for example, \cite{hu97}. To date, detections of
$E$-mode CMB polarization have been reported by DASI
\citep{kovac02,leitch05}, CBI \citep{readhead04}, CAPMAP
\citep{barkats05}, BOOMERanG \citep{montroy06}, WMAP \citep{page07},
and MAXIPOL \citep{wu_jh07}.

The White Mountain Polarimeter (WMPol) is a ground-based telescope
and receiver system designed to observe CMB polarization. This
article describes the WMPol instrument and the observing site,
including systems used to control WMPol remotely and monitor the
weather. Also presented is a summary of the data collected during
2004 and the data analysis methods and results.

\section{WMPol Instrument}

The WMPol instrument, consisting of a microwave telescope and
receiver, is located at an altitude of 3880 meters in the White
Mountains east of Bishop, California at longitude 118\degr
14\arcmin 19\arcsec W, latitude 37\degr 35\arcmin 21\arcsec N. The
University of California White Mountain Research Station (WMRS)
operates the Barcroft Facility, which includes the Barcroft
Observatory where the telescope is located. WMPol was installed in
the observatory in September 2003 and a concerted data taking
effort ran from April to October 2004.

\subsection{Telescope}

\begin{figure}
\plotone{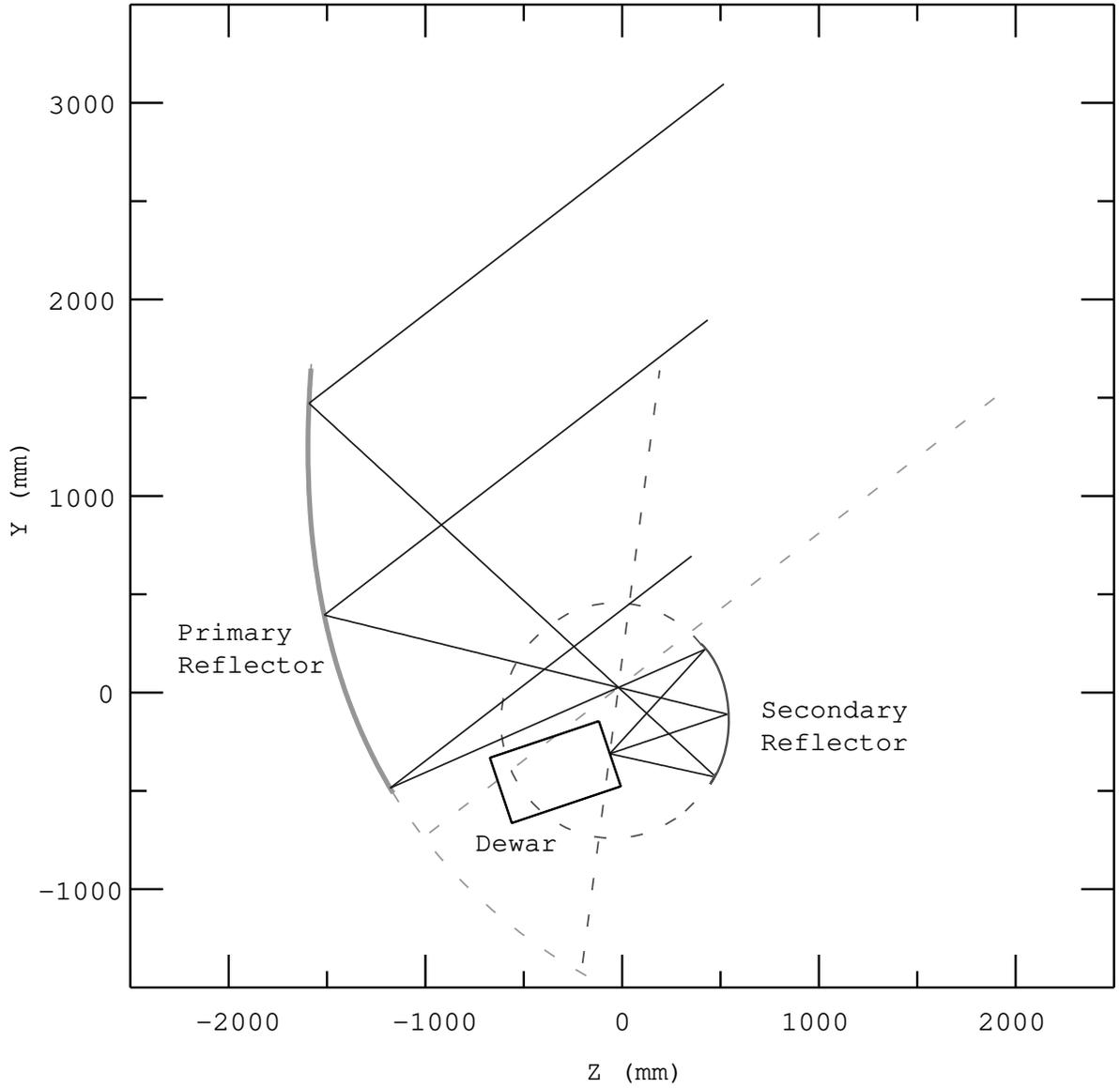}\caption{Optical design of WMPol telescope. The
primary reflector (with its parent parabola), the secondary
reflector (with its parent ellipsoid), and the dewar are shown.
Radiation from the sky is focused by the primary at the common
focal point of the primary and secondary (the origin of the plot)
and then focused by the secondary at the phase center of the
central horn in the dewar.\label{od}}
\end{figure}

The WMPol telescope is an off-axis Gregorian telescope with a
2.2-meter diameter parabolic primary. The telescope design, which
obeys the Dragone-Mizugutch condition \citep{Dragone78,Mizugutch76},
is similar to that of BEAST, a telescope dedicated to mapping CMB
temperature anisotropies \citep{childers05, figueiredo05,
meinhold05, mejia05, odwyer05, donzelli06}, and uses identical
aluminum coated carbon fiber reflectors. Figure~\ref{od} shows the
optical design including the primary, 0.9-meter diameter ellipsoidal
secondary, and the dewar that houses the receivers.

The WMPol telescope is mounted on top of a table that rotates in
azimuth. The table is attached to a shallow cone that rests on
four smaller conical roller bearings. A motor drives the motion in
azimuth by rotating one of the roller bearings through a harmonic
reducing gear.\footnote{Harmonic Drive LLC, Peabody, MA 01960,
http://www.harmonicdrive.net/} This table, as described in
\cite{Meinhold93}, was used previously to observe CMB temperature
anisotropies from the South Pole.

\begin{figure}
\plotone{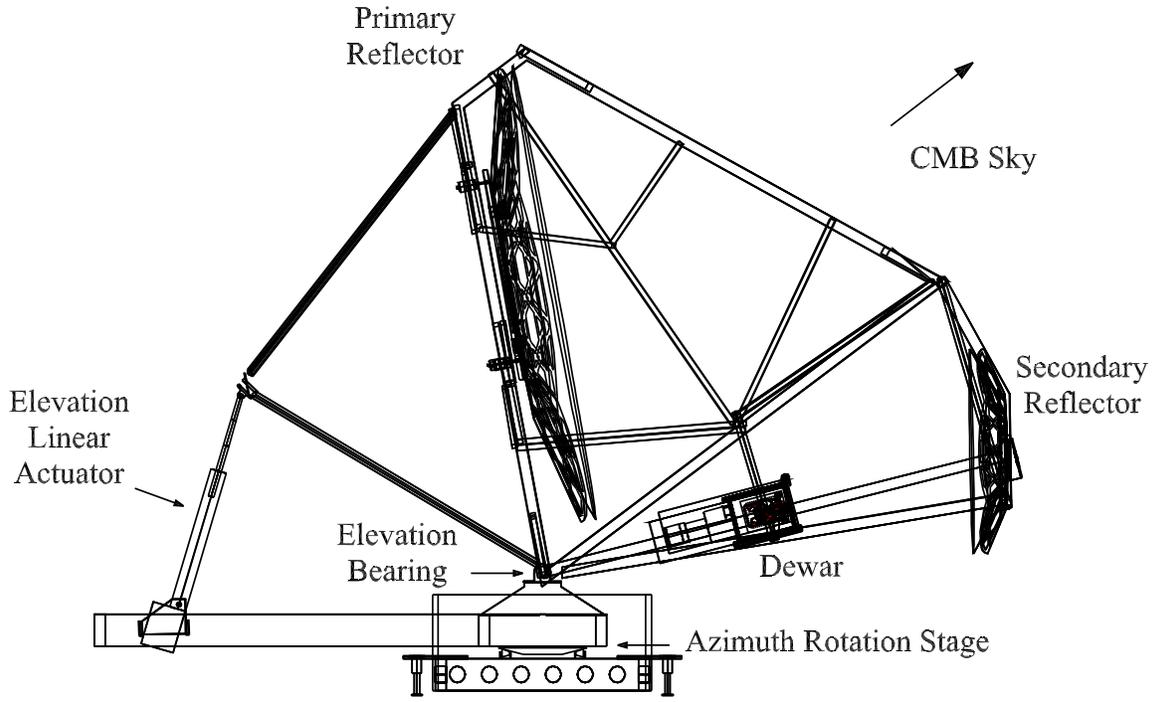} \caption{Drawing of the WMPol telescope with
primary, secondary, dewar, and frame shown.\label{te}}
\end{figure}

The telescope itself consists of a section that is mounted directly
to the table and another section (held by two bearings) that can
move in elevation via a linear actuator. The motors for the azimuth
drive and elevation actuator are controlled by a dual-axis motion
controller\footnote{Galil Motion Control, Rocklin, California 95765,
http://www.galilmc.com/} that uses relative encoders for feedback
and linear servo amplifiers\footnote{Western Servo Design, Carson
City, NV 89706, http://www.wsdi.com/} to drive the motors. Removable
switches that limit the motion in azimuth to approximately
$\pm5\degr$ were installed to prevent the possibility of damage to
the telescope during remote operation. Figure~\ref{te} shows the
major elements of the telescope.

The position of the telescope in azimuth and elevation is measured
by two 16-bit absolute encoders\footnote{Gurley Precision
Instruments, Troy, New York 12180, http://www.gurley.com/} and tilt
is measured by two biaxial clinometers,\footnote{Applied
Geomechanics Inc., Santa Cruz, CA 95062,
http://www.geomechanics.com} one with a $\pm10\arcmin$ range and the
other with a $\pm10\degr$ range. In practice, tilt of the telescope
can be controlled to within $\sim\pm 2\arcmin$ over the full
$360\degr$ azimuth range of motion.

A PC operates the telescope using code written at UCSB. This code
controls the telescope motion and reads position and tilt, takes
data through a data acquisition card,\footnote{Measurement
Computing Corp., Middleboro, MA 02346,
http://www.measurementcomputing.com/} operates an optical CCD
camera, and controls the calibrator as described in
Section~\ref{ssec:cal} below.  Temperature at two points on each
reflector is measured using AD590\footnote{Analog Devices Inc.,
Norwood, MA 02062-9106, http://www.analog.com} ambient temperature
sensors.

A great deal of effort went into arranging the power connections
to the telescope to suppress noise. Separate isolated linear power
supplies\footnote{Power-One, Camarillo, CA 93012,
http://www.power-one.com/} are used for the receiver and
housekeeping electronics and isolation transformers are used on
the computer and receiver power supplies. All telescope power
comes from a dedicated ferroresonant uninterruptible power supply,
or FERRUPS,\footnote{Powerware, Eaton Corp., Raleigh, NC 27615,
http://www.powerware.com/} and separate cables from the FERRUPS
bring power to the telescope drive, telescope computer, and
receiver. In addition, the linear servo amplifiers are installed
on the azimuth and elevation drives specifically to reduce noise.

\subsection{Receiver}

\begin{figure}
\plottwo{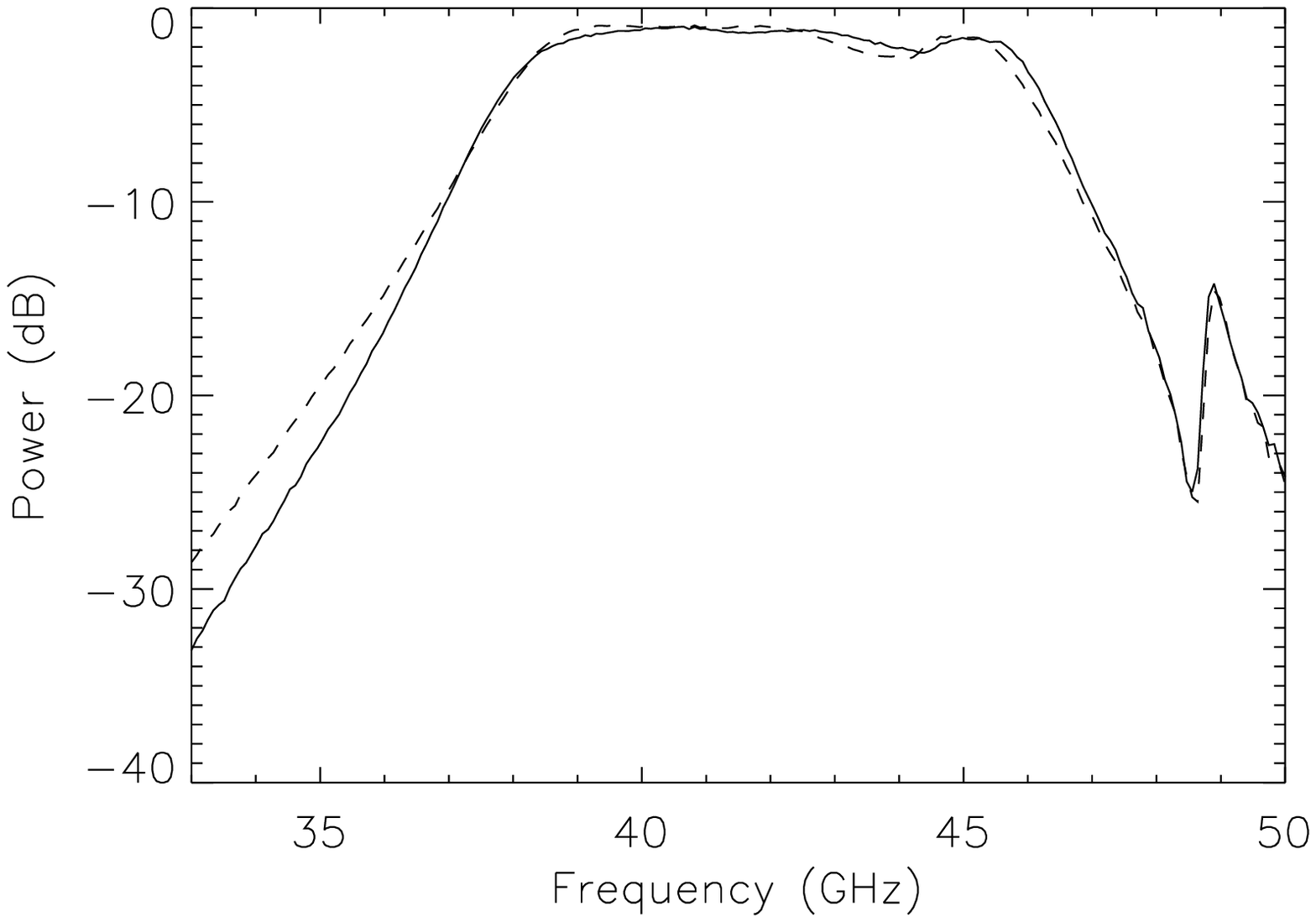}{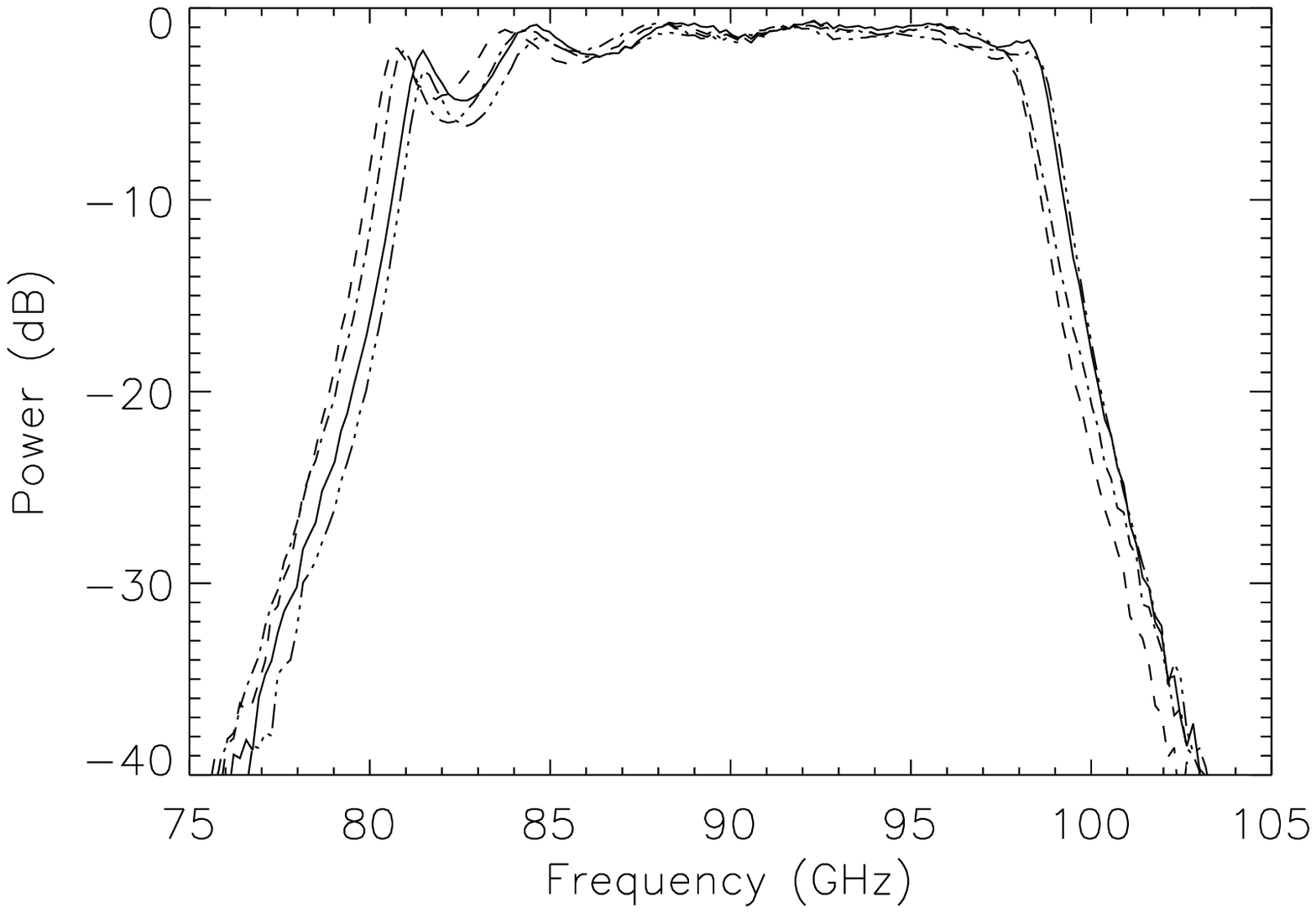} \caption{Q-band (left) and W-band (right)
filter bands. The data show transmission (normalized to input
signal) through the filters. The different lines denote each of the
filters used in the radiometers. \label{fil}}
\end{figure}

The microwave receiver system consists of two continuous comparison
polarimeters, one in Q-band ($38-46$ GHz) and one in W-band ($82-98$
GHz) plus a W-band radiometer to measure CMB temperature
anisotropies. Figure~\ref{fil} shows the Q-band and W-band filter
bands. The front ends of the receivers are contained inside an
evacuated vacuum vessel (dewar) and cooled to below 30 K with a
two-stage Gifford-McMahon cycle cryogenic
refrigerator\footnote{Leybold Vacuum Products Inc., Export PA 15632,
http://www.oerlikon.com/leyboldvacuum/} using helium as the
refrigerant. The receivers view the optics through a 10.8 cm
diameter $50\,\mu$m thick mylar vacuum window which was measured to
have a radiometric temperature of 3 K in W-band. A 0.3 cm thick
sheet of microwave transparent extruded polystyrene (``blue foam'')
is attached to the radiation shield inside the dewar between the
receivers and the window to block infrared radiation and thus reduce
the heat load on the cold stage.

The receiver topology is similar to the design of the NASA WMAP
radiometers \citep{jarosik03} and the baseline design for the Low
Frequency Instrument (LFI) for ESA-Planck \citep{seiffert02}. The
polarimeters are designed to measure the (local - i.e. in the
telescope reference frame) $Q$ Stokes parameter, or $\langle
E^2_{hor}\rangle - \langle E^2_{ver}\rangle$, while the anisotropy
measuring radiometer is designed to measure the temperature
difference between two points on the sky separated by $\sim
1^{\circ}$.

\begin{figure}
\plotone{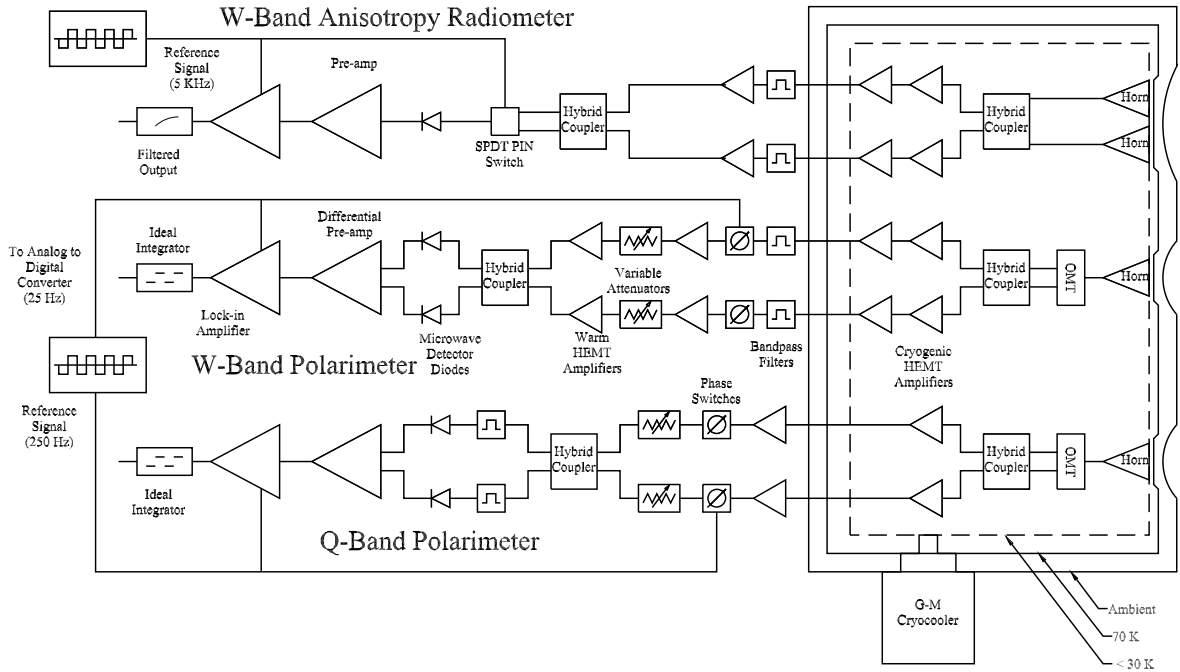} \caption{Schematic of WMPol receiver array.
Microwave radiation enters the radiometer through the vacuum
window in the dewar (right) where it is fed into corrugated scalar
feed horns. \label{ds}}
\end{figure}

\begin{figure}
\plottwo{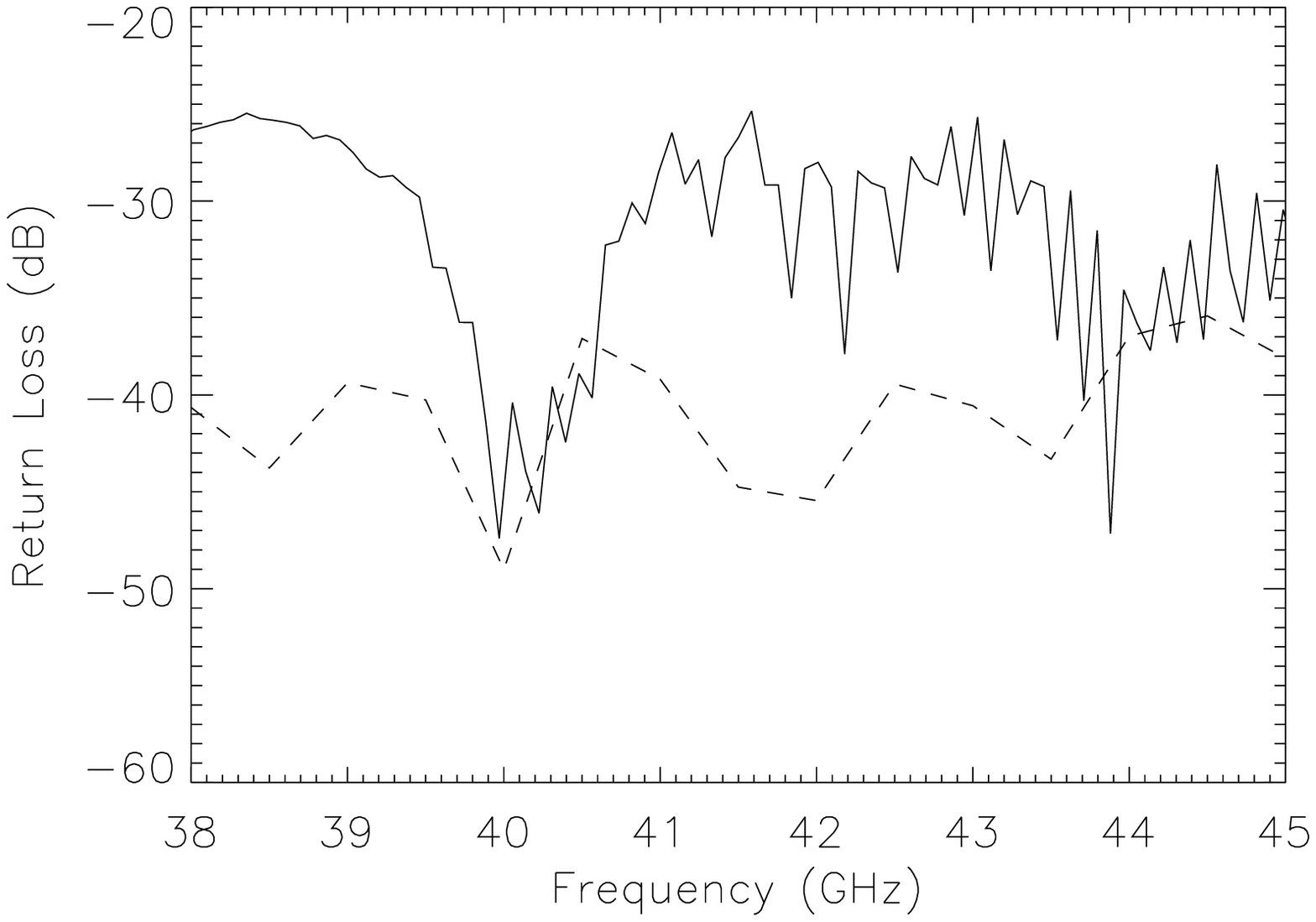}{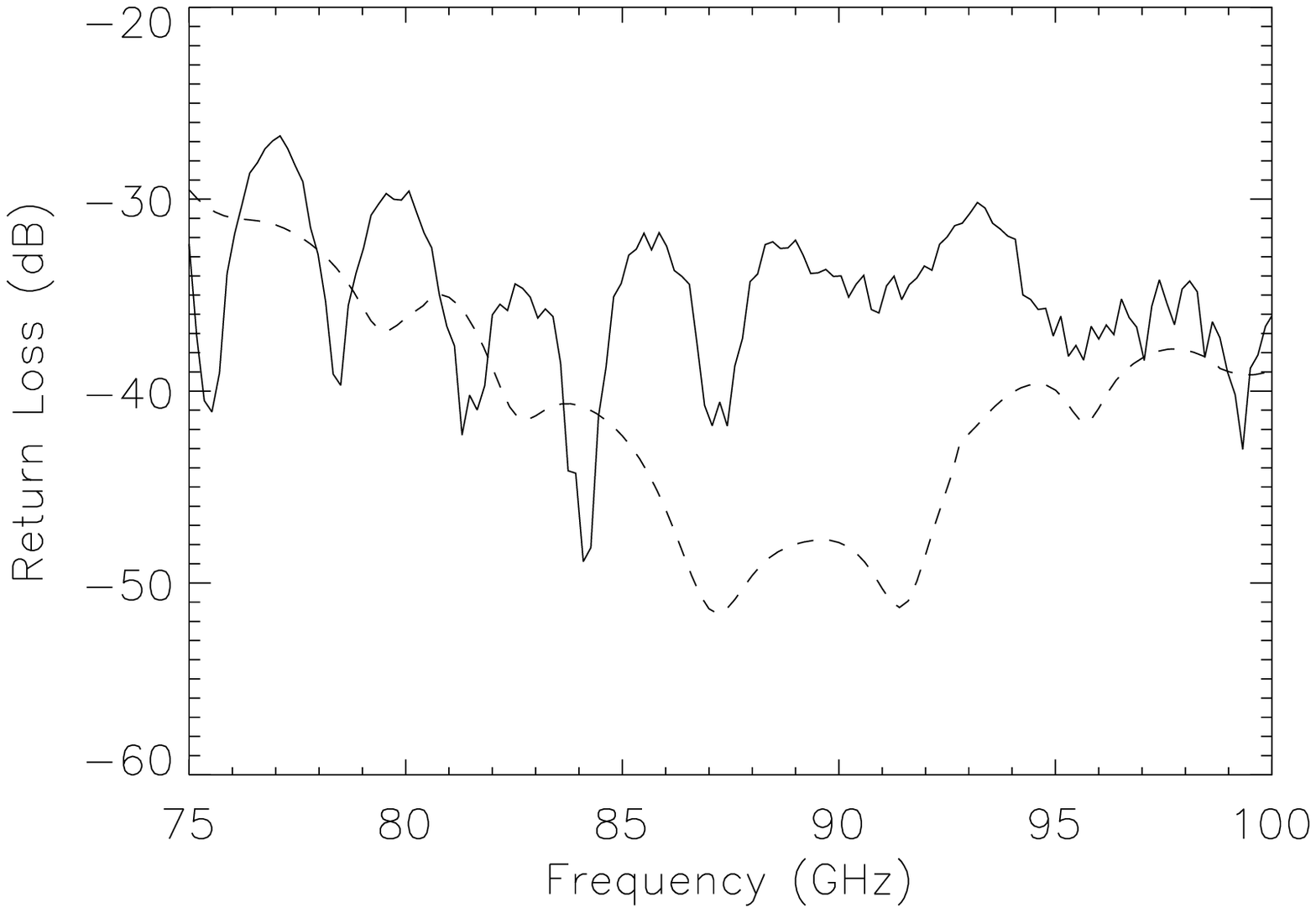} \caption{Q-band (left) and W-band
(right) feed horn return loss as a function of frequency.  Both
modeled return loss (dashed lines) and measured performance (solid
lines) are reported. \label{rl}}
\end{figure}

As shown in Figure~\ref{ds}, conical corrugated scalar feed horns
\citep{villa97, villa98} couple the microwave radiation from the
telescope to the polarimeters. As is the case for other CMB
polarization measurements, WMPol calls for feed horns with a
highly symmetric radiation pattern, extremely low
cross-polarization, low loss over a relatively large bandwidth
($\sim 20\%$ in our case), and optimal sidelobe rejection. The
optimized electromagnetic design for the W-band feeds required
throat section grooves $\sim 0.13$ mm thick and $\sim 1.5$ mm
deep. To meet these requirements, the W-band feeds were
manufactured via electroforming, with a measured deviation from
the theoretical mandrel dimensions of less than 10 microns. The
larger physical size of the Q-band feeds and their corrugations
made it possible to machine the horns from an aluminum cylinder
rather than electroforming. The main design parameters and
measured performances for the WMPol feeds are shown in
Table~\ref{t1}. Figure~\ref{rl} displays the return loss of the
Q-band and W-band feeds.

\begin{figure}
\plotone{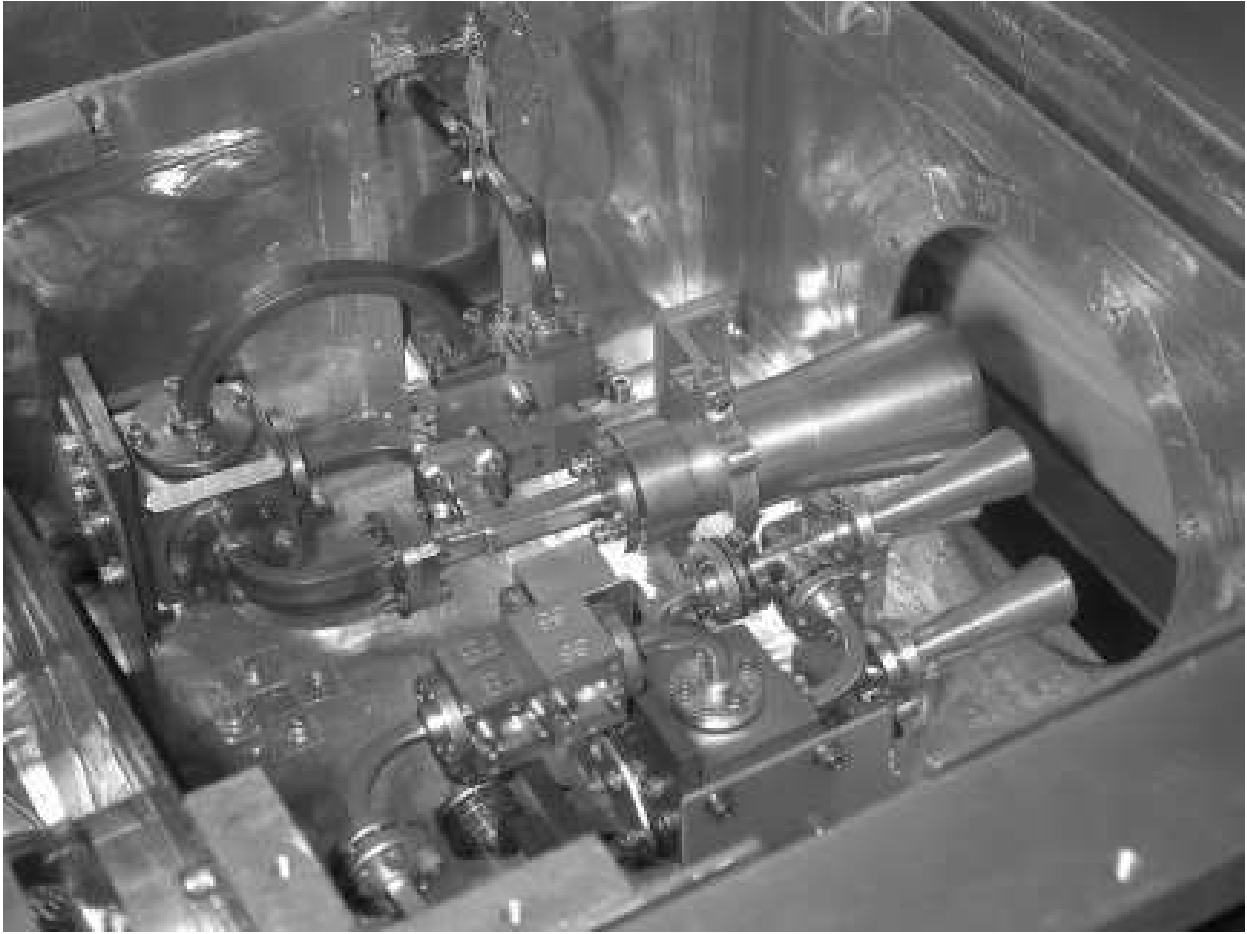} \caption{View inside the dewar. The W-band
polarimeter horn (top, small horn) is aligned with the optical
axis of the telescope and the two W-band horns underneath are for
the W-band anisotropy radiometer. The Q-band polarimeter horn is
tilted $4\degr$ towards the W-band polarimeter horn due to the
curvature of the focal plane when off of the optical axis.
\label{dp}}
\end{figure}

For each polarimeter, the radiation from a horn feeds into an
orthomode transducer (OMT). The OMTs use an asymmetric design with
specifications as reported in Table~\ref{t2}. From the OMT, the
two polarizations are split and go into the two inputs of a $180$
degree hybrid coupler. The two outputs of the $180$ degree hybrid
lead to cryogenic amplification with High Electron Mobility
Transistor (HEMT) monolithic microwave integrated circuit (MMIC)
amplifiers and then more amplification outside of the dewar, plus
bandpass filtering in the case of the W-band polarimeter, before
leading to the second 180 degree hybrid. The path between the two
hybrid couplers is phase-sensitive and $180\degr$ phase switches
are installed to facilitate changing one of the path lengths by
$\lambda/2$ at a modulation frequency of $100 - 250$ Hz.
Adjustable attenuators are also installed in the phase-sensitive
paths to match the gain of the two paths and thin brass shims
(approximately $\lambda/40$ or 0.23 mm for Q-band and 0.10 mm for
W-band) are used to match the phase lengths of the two paths. The
outputs of the second hybrid go to square-law detector diodes. The
bandpass filters for the Q-band receiver are placed between the
hybrid and the detector diodes. The diode outputs are fed into a
$\times100$ gain differential amplifier before going to a lock-in
amplifier referenced to the phase modulation frequency. The
purpose of differencing the detector diodes is to improve the
1/$f$ noise performance of the instrument. Finally, the output of
the lock-in goes to an ideal integrator and the data acquisition
card in the telescope computer. The individual diode signal levels
after $\times100$ gain are also recorded. These DC levels do not
benefit from the improved 1/$f$ performance that differencing
provides and do not go to a lock-in amplifier. However, these
measurements were found to be very helpful for calibration and
measuring sky temperature as described further below.
Table~\ref{t3} identifies the sources of various elements of the
polarimeters and Figure~\ref{dp} is a photograph of the inside of
the dewar.

The W-band CMB temperature anisotropy radiometer is similar to the
polarimeters except that two horns feed into the first hybrid
coupler and the outputs of the second hybrid coupler feed into a
single pole double throw PIN switch modulated at 5 kHz which then
feeds a square law detector diode. Phase sensitive detection at
this modulation frequency provides a signal that is proportional
to the temperature difference between the two points on the sky
that the horns view. Due to space limitations and other
considerations, adjustable attenuators are not used in the phase
sensitive paths; instead, the gains are matched through small
changes in the bias voltages.

All the radiometer electronics and the microwave components
outside of the dewar are contained in a contiguous aluminum
shield. To reduce noise, the dewar is electrically isolated from
the telescope frame except for two ground connections. One ground
connection goes from the dewar to the cryogenic refrigerator cold
head and the helium compressor through the compressed gas lines
and electrical cable that connect the cold head to the compressor.
Another ground connection from the dewar to the radiometer power
supply ground ensures that the dewar ground connection is not
severed if the cryocooler power cable is disconnected from the
wall.

The noise temperatures of the Q-band and W-band polarimeters are
measured using manual calibrations of the non-differenced DC
channels. The resulting Q-band noise temperature is 127 K and the
W-band noise temperature is 120 K. The high noise temperature,
especially for the Q-band system, is mostly due to the quality of
amplifiers that were available to us at the time that the
receivers were assembled, tested, and deployed. As the measured
room temperature losses due to the OMT and hybrid coupler before
the cryogenic amplifiers are less than 0.5 dB for Q-band and 1.5
dB for W-band, the maximum contribution to the polarimeter noise
temperature due to front end losses, assuming the polarimeters are
at a physical temperature of 30 K, is 4 K for Q-band and 12 K for
W-band. The W-band anisotropy measuring radiometer failed soon
after the telescope was installed in the Barcroft Obsevatory and
repeated attempts to discover the problem have not been
successful. Table~\ref{t4} is a list of receiver noise
temperatures, effective bandwidths, and sensitivities for the Q
and W polarimeters.

During the testing and commissioning of the instrument, it was
found that the polarized offsets were rather high ($\sim26$ K)
which limited the amount of gain that the polarimeter output could
be multiplied by before going to a data acquisition channel. The
offset is due to a change in DC signal out of the diodes when the
phase switches are in different states. We believe the origin of
the offset is the asymmetry between the two output arms of the
OMT, which is not correctable in our radiometer design. A circuit
was designed and successfully implemented to remove the offset
before sending the signal to the data acquisition system.

\section{The Barcroft Observatory}

Although microwave atmospheric emission is known to be not
significantly polarized \citep{hanany03}, the choice of a suitable
observing site is mandatory to optimize the data taking
efficiency, reduce systematic effects and avoid large baseline
drifts. The WMRS Barcroft Facility provides an excellent site for
CMB observations. The addition of a remote control system to
operate WMPol and weather monitoring at the Barcroft Observatory
improved our ability to collect data safely and efficiently.

\subsection{Sky Temperature}

\citet{marvil06} and a previous study of atmospheric emission at
Barcroft \citep{bersanelli95} show low water vapor content at the
Barcroft Facility, with observed limits as low as $\sim 0.2$ mm.
Assuming precipitable water vapor content in the range $0.2-6$ mm
and calibrating the atmospheric emission model of \citet{danese89}
with more recent data, we expect zenith atmospheric (sky)
temperatures at the centers of the WMPol bands in the ranges of
$9-12$ K at 43 GHz and $7-16$ K at 90 GHz. It should be noted that
the atmospheric emission spectrum in the $40-45$ GHz range,
dominated by O$_2$ broadened line complex, is extremely steep so
that the tail of the band can contribute significantly to the
zenith sky temperature (see Figure~\ref{fil} for the WMPol Q-band
filter band).

The sky temperature is measured with sky dips. For WMPol, the sky
dips are scans in elevation from $32\degr$ to $48\degr$ at azimuth
$0\degr$, which give the zenith sky temperature after fitting to
the model
\begin{equation}
T_{\mathrm{antenna}} = T_{\mathrm{system}} +
\frac{T_{\mathrm{zenith}}}{\sin(\theta)}
\end{equation} where $\theta$ is the elevation angle. During the
observing campaign, the measured mean zenith sky temperature was
$(9 \pm 0.2)$ K and $(10 \pm 0.6)$ K for Q-band and W-band
respectively with measurements ranging from $8-11$ K for Q-band
and $9-15$ K for W-band. These values agree well with the
predicted values and measurements from \citet{marvil06},
\citet{childers05}, and \citet{bersanelli95}.

\subsection{Remote Control System}

We considered it quite advantageous to be able to operate the
telescope remotely. In order to facilitate remote operation and
data collection a number of elements were installed in the
observatory to enable monitoring and control of the telescope.

The core of the remote control system is a Stargate automation
system.\footnote{JDS Technologies, Poway, CA 92064-6876,
http://www.jdstechnologies.com} The Stargate has the capability to
control X10 (i.e. on/off signals sent over power lines) devices
and on-board mechanical relays as well as read digital and analog
inputs. The Stargate can accept commands by phone line, internet,
or computer via a serial port. The X10 and mechanical relays are
used to turn on and off (or power cycle to reset, as needed) the
computers and other elements of the telescope.

Another key element of the remote control system is the ability to
communicate with and operate the computers in the observatory from
other locations. This is achieved by using the Remote
Administrator (Radmin) software.\footnote{Famatech,
http://www.radmin.com/} Radmin allows for control of the computers
at the observatory from the Barcroft station and also from Santa
Barbara and potentially anywhere with an internet connection.

For remote operation, it is important to be able to monitor the
weather so that the telescope may be protected by closing the
observatory shutter in case of high winds or precipitation. As
described in Section~\ref{ssec:weather}, a weather station and a web
camera (pointed towards north) are used to monitor sky and weather
conditions. The weather station data and camera images are archived
every ten minutes and downloaded to a web site at least every half
hour for easy monitoring. We also have the ability to view the web
camera live.

The optical CCD camera\footnote{Watec America Corp., Las Vegas, NV
89120} installed on the telescope to assist with pointing
reconstruction can be used to monitor cloud conditions at night by
checking whether or not Polaris is visible. The CCD camera is also
sensitive to clouds during the day.

While it is critical for remote operations to maintain
communication with the observatory, the location of the Barcroft
Facility and potential for rough weather makes communication a
challenge. WMRS has installed two systems to connect to the
internet, a satellite connection\footnote{StarBand Communications
Inc., McLean, VA 22102, http://www.starband.com/} and a wireless
T1 connection through radios\footnote{Wi-LAN Inc., Calgary, AB
Canada T1Y 7K7, http://www.wi-lan.com/} at the observatory, White
Mountain peak (longitude 118\degr 15\arcmin 17\arcsec W, latitude
37\degr 38\arcmin 04\arcsec N, altitude 4342 meters), and the WMRS
Owens Valley Lab, Bishop, CA. We also installed a cell phone base
station with our cell phone line connected to the Stargate system.
During the first campaign of data taking, lightning and other
extreme weather conditions caused some communications outages but,
fortunately, no damage to the telescope was suffered during these
incidents. As more experience is gained with the communications
infrastructure, it is expected to be more reliable.

Another potential issue is power outages due to inclement weather.
All the critical elements in the observatory are on FERRUPS for
protection and short term power. A planned upgrade to put the
shutter control on a UPS and have the shutter close automatically if
there is a loss of power would resolve this issue.

\subsection{Weather Monitoring}\label{ssec:weather}

To help protect the WMPol telescope from inclement weather and in
order to aid data selection, a commercially available weather
station\footnote{Davis Instruments Corp., Hayward, CA 94545,
http://www.davisnet.com/} and USB web camera\footnote{Creative Labs
Inc., Milpitas, CA 95035, http://us.creative.com/} were installed at
the Barcroft Observatory. The weather station has the ability to
measure temperature, relative humidity, barometric pressure, wind
speed and direction, and solar irradiance. Relative humidity in
conjunction with the web camera during the day and a CCD camera used
for imaging of stars during the night are used to determine the
onset of bad weather. To avoid damage to the telescope, the shutter
was closed and data taking suspended if inclement weather was
imminent or if the average wind speed exceeded 20 mph (8.9
m$\cdot\mathrm{s}^{-1}$) and/or wind gusts exceeded 30 mph (13
m$\cdot\mathrm{s}^{-1}$).

During the April through October 2004 observing campaign, the mean
temperature was $3.3\degr$C with an average daily temperature swing
of approximately $8\degr$C and the wind speed was lower than our
damage-avoidance threshold 93 percent of the time. \citet{marvil06}
give a more detailed analysis of the Barcroft Observatory weather
station data

The web camera images were obtained every 10 minutes and were rated
by eye. During the observing campaign, the sky was ``Clear,''
corresponding to a cloudless period, $46\%$ of the time and ``Partly
Cloudy,'' indicating some scattered clouds, $30\%$ of the time.
Periods that were rated ``Mostly Cloudy'' ($15\%$ of the time) or
``Overcast'' ($9\%$ of the time) were considered to be times when
the weather conditions would not result in useful radiometer data.

\section{Observing Campaign}

After the telescope was installed in the Barcroft Observatory,
Winter 2003-04 was spent commissioning and testing the telescope. We
define 2004 April 23 as the official beginning of the observing
campaign which ended on 2004 October 17 with the first winter storm
of the year and the scheduled seasonal closure of the Barcroft
Facility.

\subsection{Data Acquisition}

During the 178-day campaign, data were recorded during 2169 hours,
or just over half of the available time. Table~\ref{t5} describes
the observing time lost due to various issues. Inclement weather,
including precipitation and high wind speed, was a major factor
and the Stargate automation system was damaged on multiple
occasions, we believe due to lightning, causing a great deal of
down time. Several levels of surge protection were added to the
Stargate during the observing campaign but the seasonal lightning
storms ended before it could be determined whether this solved the
problem.

The cryogenic refrigerator used to cool the front ends of the
radiometers also halted data collection for two reasons. First,
the cold end of the refrigerator is known to have a mechanical
problem that slightly reduces performance and seems to cause the
cooler and dewar to warm up on its own after about fourteen days
of continuous operation. In order to recycle the system, the
inside dewar temperature has to be brought up all the way to
ambient temperature ($\gtrsim 280$ K) before cooling down again;
this process requires approximately 24 hours. The other reason for
data loss was overheating of the refrigerator's helium compressor
during the middle of the day during summer. The compressor is
liquid-cooled with a fan plus radiator to draw heat away from the
compressor motor. A thermal cutout prevents damage due to
overheating. During winter, a $50-50$ water-antifreeze (ethylene
glycol) mix is used to prevent the coolant from freezing during
times when the compressor is shut down. This mix has reduced
thermal properties as compared to pure water and is not sufficient
to transfer heat during midday from the compressor motor in the
increased summer temperature conditions. Once this issue was
discovered, replacing the anti-freeze mix with distilled water
solved the problem.

\subsection{Pointing Reconstruction}

In order to determine our pointing, we used the Moon and stars to
align the optical CCD camera (approximate field of view of $2\fdg7
\times 2\fdg0$) with the radiometric channels. The size of each CCD
camera image pixel, $15\arcsec \pm 1\arcsec$, was determined by
observing known stars fields. Observations of the Moon were used to
find the central pixel positions of the radiometers in the CCD
camera. Subsequent observations of the pixel position of Polaris in
the CCD camera allowed us to calculate any offset between the
polarimeters and the position of the telescope as defined by the
encoders. We estimate our pointing error to be $\pm3\arcmin$ for the
Q-band polarimeter and $\pm2\arcmin$ for the W-band polarimeter.
This pointing error is a small fraction of the beam size. To verify
that there were no changes in the pointing on a day-to-day basis,
CCD camera images were recorded at all times at a rate around one
per minute. In addition we spent 1200 seconds each night of
observation staring at Polaris and $\lambda$ Ursae Minoris at
approximately 7:00 UT to build up a data base to reveal any changes
in the pointing of the telescope or CCD camera.

\subsection{Observing Strategy}

WMPol used a constant elevation scan strategy, similar to the
strategies employed by PIQUE \citep{hedman01, hedman02}, COMPASS
\citep{farese03, farese04}, and CAPMAP \citep{barkats05}. The
telescope maintained an elevation of $37\fdg6$ while scanning the
sky in azimuth $\pm45\arcmin$ about the North Celestial Pole (NCP)
with a period of 9.5 seconds. An advantage of this strategy is
that the atmospheric contribution remains nearly constant over
short time intervals. Sky rotation allows the telescope to observe
a circular region on the sky centered on the NCP. The useful
region of the sky observed has an area of 5.5 $\mathrm{deg}^2$ in
the Q-band polarimeter channel and 1.1 $\mathrm{deg}^2$ for the
W-band polarimeter channel. The larger coverage area of the Q-band
polarimeter is due to a $54\arcmin \pm 1\arcmin$ offset of the
Q-band beam from the optical axis, the value of which was
determined by observations of the Moon with both polarimeters.

WMPol only measures the Stokes parameter $Q'$ in the frame defined
by the orthomode transducer of the instrument. The sky Stokes
parameters $Q$ and $U$, expressed according to IAU convention
\citep{iau74}, are related to WMPol measurements by
\begin{equation}\label{e:QUtransform}
 Q'\propto Q\cos2\psi+U\sin2\psi,
\end{equation}
where $\psi$ is the parallactic angle measured from North and
increasing through East (i.e. the angle at the intersection between
two great circles passing through the observation point, one
containing the NCP and the other the WMPol zenith). Due to the WMPol
scan strategy, $\psi$ remains constant and approximately equal zero
during data acquisition. So, in the absence of noise, Equation
(\ref{e:QUtransform}) allows us to interpret $Q'\simeq Q$. Due to
the choice of this scan strategy, WMPol measurements of $U$ Stokes
parameter are negligible and are not considered during further
analysis.

\subsection{Beam Determination}

\begin{figure}
\plotone{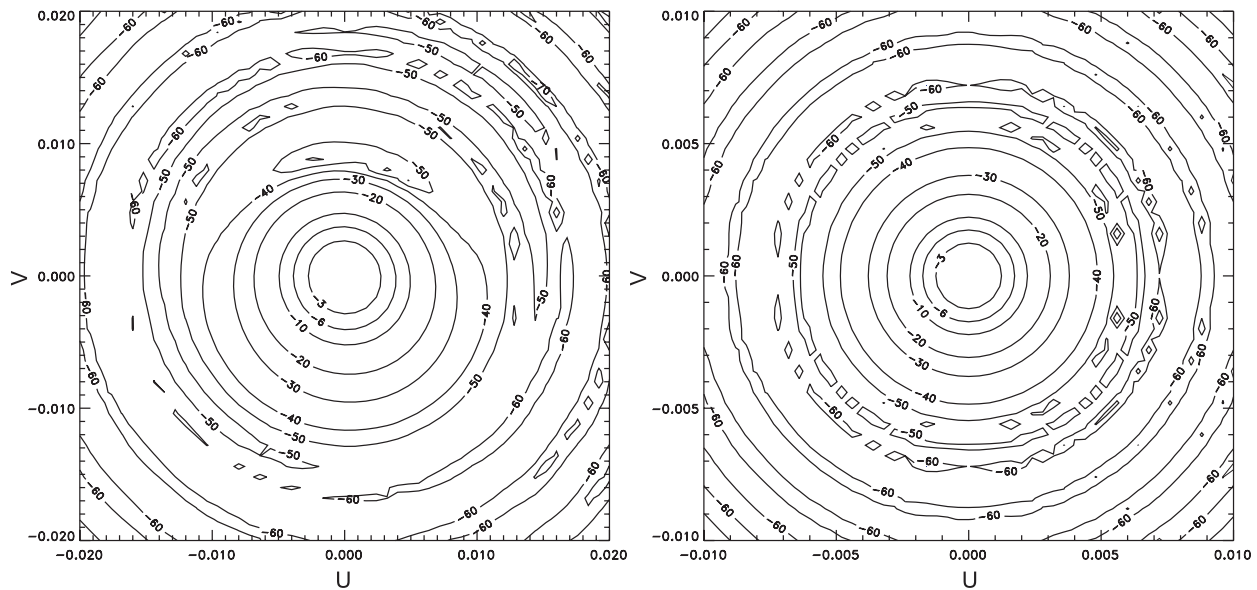}\caption{Contour plots of beam models for the Q-band
(left) and W-band (right) polarimeters. The axes are dimensionless
direction cosines given by $u=\sin\theta\cos\phi$ and
$v=\sin\theta\sin\phi$, where $\theta$ and $\phi$ are spherical
coordinate angles with respect to the maximum beam response.
However, $u$ and $v$ are well approximated by radians for such small
angles. The first three contours correspond to -3,-6, and -10 dB
below the peak. Note the low level asymmetry (coma lobe) of the
off-axis Q-band feed. \label{bmQP}}
\end{figure}

The beam sizes of the radiometers on the sky after passing through
the optics were modeled using GRASP8.\footnote{TICRA, Copenhagan,
Denmark, http://www.ticra.com} Geometrical optics with geometrical
theory of diffraction was used for the secondary reflector, while
for the primary reflector physical optics was applied. The
geometry was analyzed in transmitting mode (i.e. from feed to
sky). In the calculations, far field effects were neglected at the
feed level as the feed aperture is only about $4.5\lambda$ and the
secondary reflector is at the far field of the feeds. Gaussian
beam models of the feeds for both Q-band and W-band were used with
modeled FWHM beam sizes of $19.4$ degrees at 41.5 GHz and $19.3$
degrees at 92.5 GHz. Each beam was calculated using a regular
(u,v) grid centered on each beam peak in the range (-0.02, 0.02)
at 41.5 GHz and (-0.01, 0.01) at 92.5 GHz, for both u and v,
corresponding to $1\fdg15\times1\fdg15$ and $0\fdg575\times
0\fdg575$ on the sky. Figure~\ref{bmQP} shows contour plots of the
beam models for the polarimeters. To characterize each beam, an
elliptical Gaussian fit of the beam in the whole angular region of
calculation was obtained. The modeled characteristics are reported
in Table~\ref{t6} with expected beam sizes (average FWHM) of the
Q-band and W-band polarimeters of $18\farcm8$ and $8\farcm5$
respectively.

\begin{figure}
\plotone{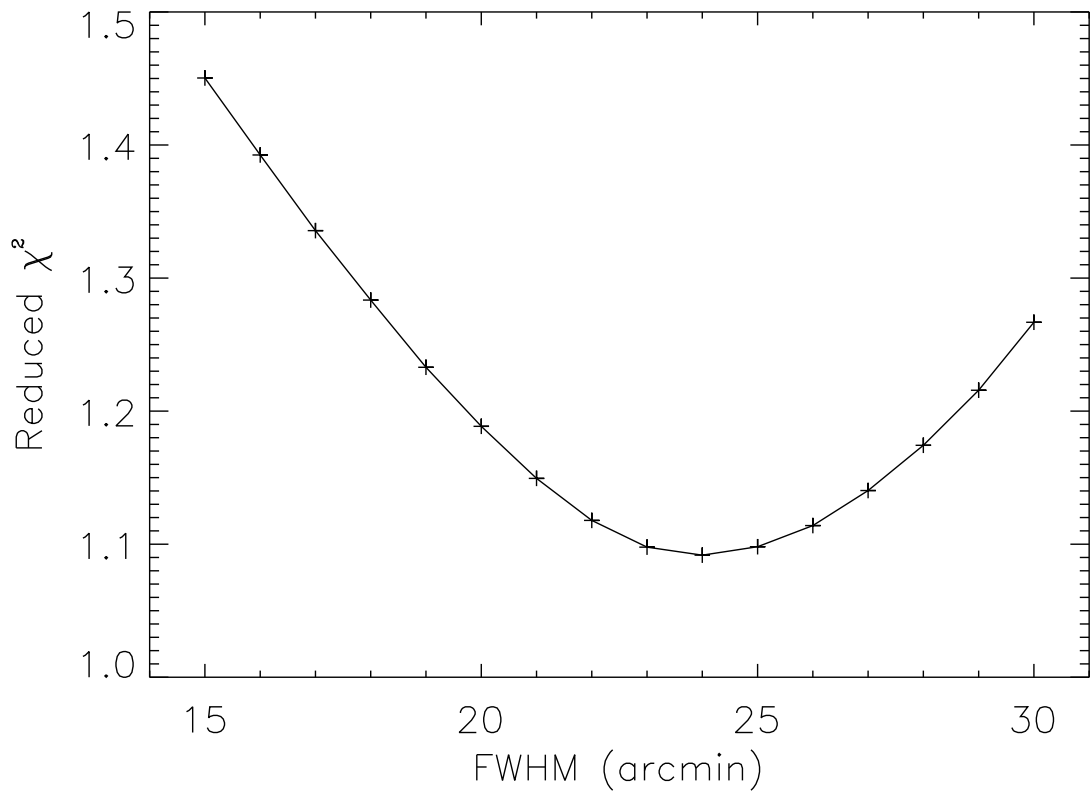}\caption{Determination of the Q-band beam size. A
least squares fit between observations of Tau~A and a point source
convolved with a Gaussian beam was used to determine the beam size.
The measured FWHM beam size of the Q-band polarimeter is $24\arcmin
\pm 3\arcmin$. The number of degrees of freedom is 347,
corresponding to 347 HEALPix pixels (3.4 arcmin resolution, nside =
1024) that cover a 1.1 square degrees circular sky area centered on
Tau~A.\label{f:beamsize_Q_chi_reduced.eps}}
\end{figure}

Measurements of the actual beam size were accomplished using the
Moon and Tau A. A least squares fit between the observations and the
source convolved with a Gaussian beam was used to determine the beam
size. The measured FWHM beam size of the Q-band polarimeter is
$24\arcmin \pm 3\arcmin$ based on observations of Tau A, as shown in
Figure \ref{f:beamsize_Q_chi_reduced.eps}. This result is similar to
the $22\arcmin \pm 2\arcmin$ quoted in \citet{childers05}.
Presumedly due to 1/$f$ issues, Tau~A was not detected with the
W-band channel but, by using Moon observations and a model for the
Moon as given by \citet{keihm84}, a FWHM beam size of $12\arcmin \pm
3\arcmin$ was found.

\subsection{Calibration}\label{ssec:cal}

Calibration of the receiver was accomplished with multiple
techniques. A manual calibration of the non-differenced channels
of the polarimeters was performed by recording the radiometer
signals while an ECCOSORB\footnote{Emerson \& Cuming Microwave
Products, Randolph, MA 02368, http://www.eccosorb.com} ambient
temperature load was placed in front of the radiometers and
comparing to the signal levels when viewing the sky.

\begin{figure}
\plottwo{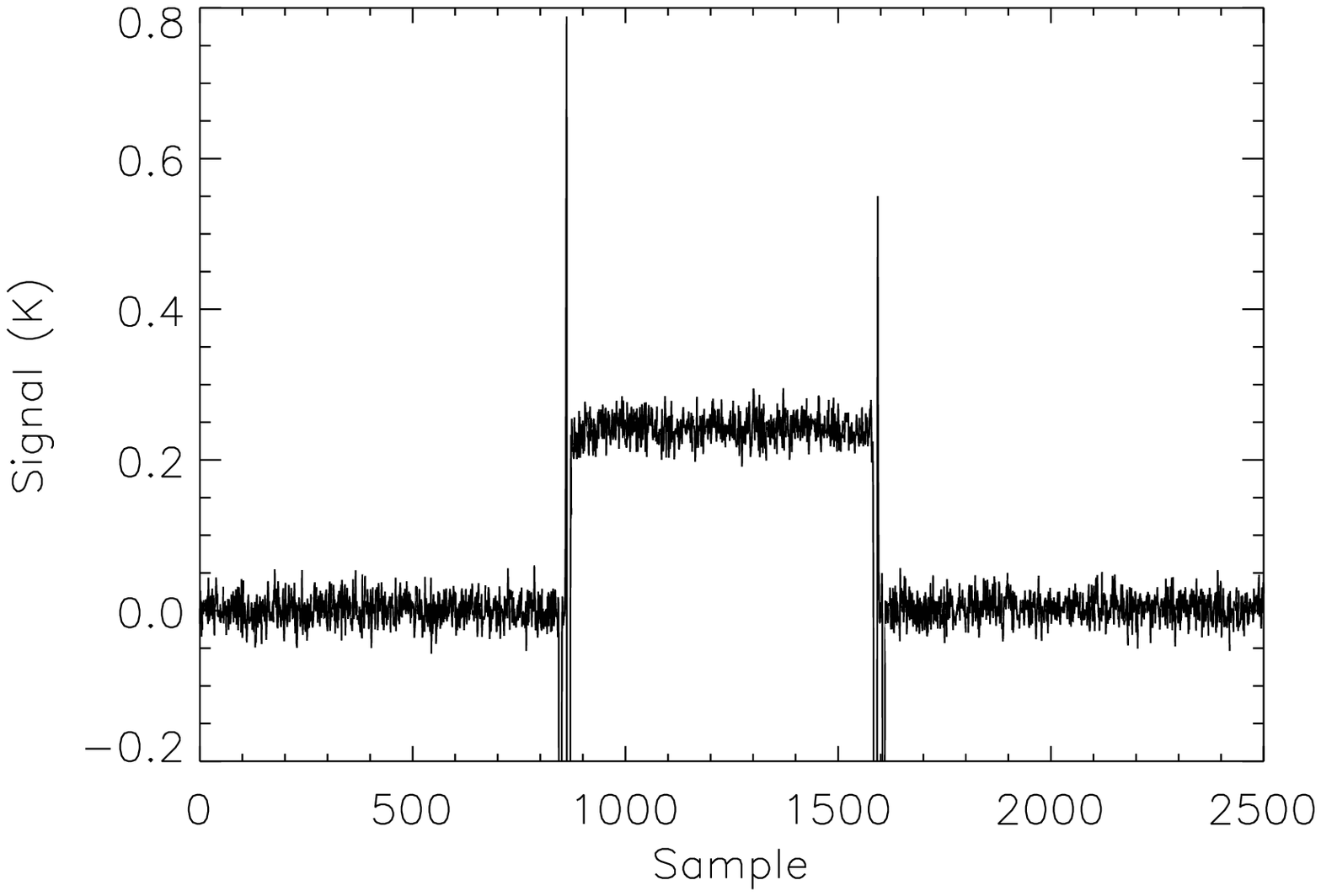}{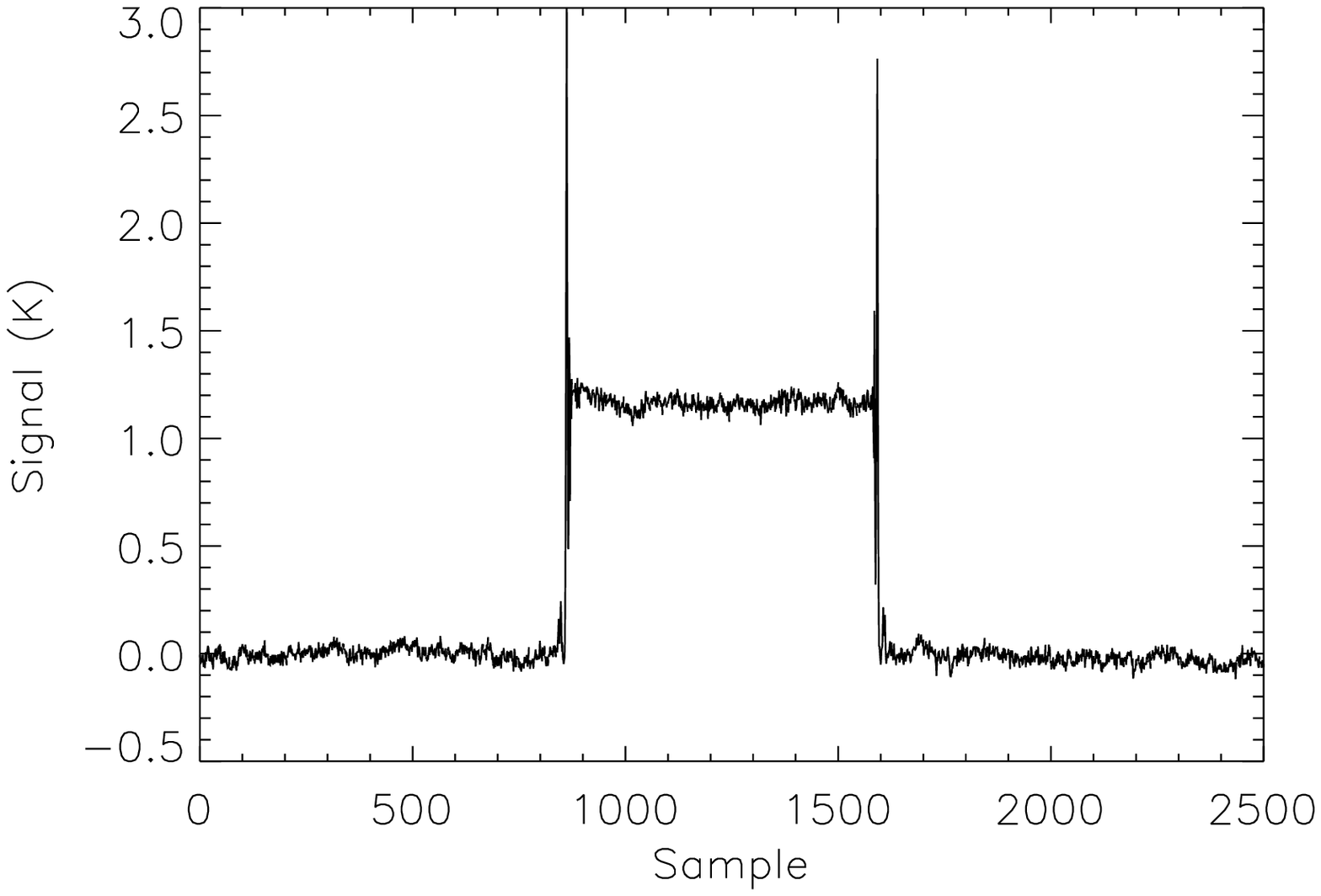} \caption{Example of automatic
calibration sequence in Q-band (left) and W-band (right), 25 Hz
data rate. Placing the dielectric film at a $45\degr$ angle in
front of the radiometers creates a relatively small polarization
signal that is easily detected. \label{auto}}
\end{figure}

\begin{figure}
\plotone{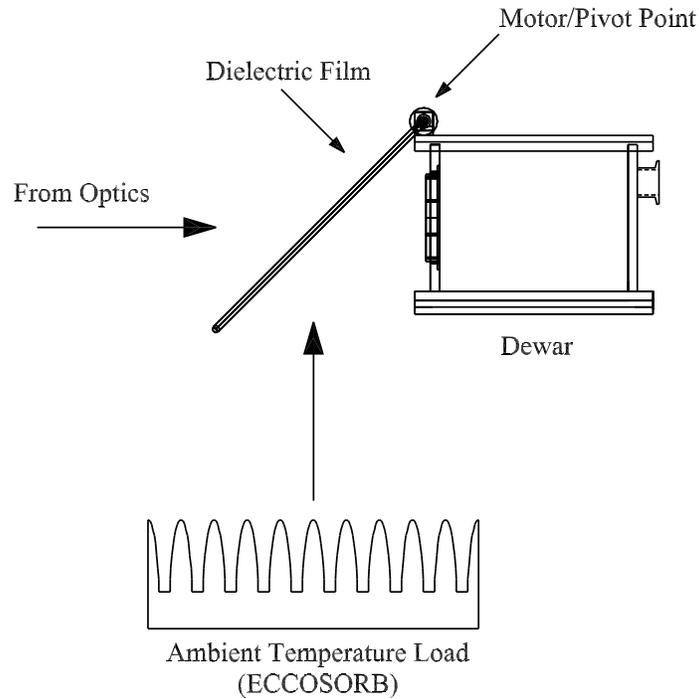} \caption{Automatic calibrator. The dielectric
sheet is supported by a frame constructed of fiberglass tubing.
During a calibration sequence, the dielectric sheet is moved in
front of the dewar window as shown and unpolarized radiation from
the ambient temperature load and the sky (through the optics) is
slightly polarized after reflecting off of or passing through the
sheet to the polarimeters. This small polarization signal is due
to the difference in the reflection coefficients of the sheet for
radiation polarized either parallel or perpendicular to the plane
of incidence. After calibration, the frame is rotated away from
the polarimeters about the pivot point until it rests on top of
the dewar. Limit switches are used to properly position the frame
and a digital flag is recorded in the data to identify when
calibration sequences occur. \label{cal}}
\end{figure}

To calibrate the polarization channels, a grid of 150 $\mu$m
diameter copper wires with a 485 $\mu$m spacing was used. This
grid was held at an angle of 45 degrees with respect to the dewar
window and provided a large polarized signal due to the difference
between the temperature of the sky and an ECCOSORB ambient
temperature load. As this calibration procedure was manual and
required changing the lock-in gain, we also implemented an
automated calibrator with a thin dielectric (polypropylene) sheet
to perform a relative calibration every 10 minutes. This technique
is similar to that described in detail in \citet{odell02}. For
each manual calibration of the polarized channels, a transfer
standard was calculated and used to convert from the automatic
calibrator. This transfer standard, $\alpha$, takes the form
\begin{equation}\label{e:transfer_standard}
\alpha = \frac{\vert \Delta V_{\mathrm{film}} \vert}{G \vert
T_{\mathrm{ref}} - T_{\mathrm{sky}} \vert}
\end{equation} where $\Delta V_{\mathrm{film}}$ is the polarimeter signal
offset due to placing the film in front of the dewar,
$T_{\mathrm{ref}}$ is the antenna temperature of the ambient
reference load, $T_{\mathrm{sky}}$ is the measured mean radiometric
sky temperature at elevation $37\fdg6$, and $G$ is the calibration
constant (volts per kelvin) from a manual grid calibration.
Figure~\ref{auto} shows examples of the signal level change in the
polarization channels due to the automatic calibration and
Figure~\ref{cal} shows the orientation of the automatic calibrator
(or copper wire grid for manual calibration) with respect to the
dewar. We estimate the calibration uncertainty on the adopted
procedure to be $20\%$.

We verified our Q-band calibration by using Tau~A as a target. This
source was observed through 7 constant elevation scans during 41
minutes on 2004 September 1. The scans were combined to obtain a
Stokes Q map of Tau~A at a parallactic angle of $58\fdg3 \pm
0\fdg4$. The map obtained was compared with Tau~A on Q-band maps
from the WMAP sky survey \citep{page07, hinshaw07}. This procedure
allowed us to verify the Q-band automatic calibration within
measurement error.

\section{Data Analysis}

Scientific and housekeeping data were stored in files corresponding
to 20 minutes in length. For each file and channel we computed
statistical estimators, polarimeter white-noise levels, the weather
condition, calibration constants, and other information useful in
classifying the data. This procedure allowed us to investigate long
time scale behavior of instrument and data, and flag data before
producing Stokes Q sky maps.

\subsection{Data Selection}

Table~\ref{t7} gives the number of hours of data that were cut from
the raw data to create a CMB data set. Data files for times when the
telescope was not scanning about the NCP were not used except where
data were specifically taken for the purposes of pointing
reconstruction, beam size determination, and calibration as
discussed below. Also, files shorter than 60 seconds were cut as
these files were not expected to contain enough data to be useful.
During midday in the late spring and summer the Sun was high enough
in the sky to shine directly on the telescope. This not only caused
the secondary reflector to heat up by as much as 20\degr C but also
caused the warm components of the polarimeters to heat up to a
temperature higher than the servo set point. Changes in signal and
gain due to this heating caused the DC signal levels of the
polarization channels to exceed the limits of the data acquisition
electronics and we labeled these data as ``saturated.''

The Q-band and W-band polarization-sensitive channels of each data
file were visually inspected. The time-ordered data before and
after removing the DC offset on these channels for each half scan
(one sweep in azimuth from $-45\arcmin$ to $+45\arcmin$ or the
reverse) was examined along with the power spectral density of
both channels. The rms of the data after removing the offset was
also calculated. Many files that were found to have a rms
significantly different from the mean had unusual features in both
polarization channels that were presumed to be due to clouds
passing through the beam. These files were removed from the data
analysis.

After the rms cut, we found that there were data that passed the
cut that were collected during bad weather (i.e. web camera image
rating equal to ``Mostly Cloudy'' or ``Overcast'') and these data
were discarded. Another small amount of data was cut due a cold
plate temperature that was too high as a result of the dewar
warming up. After cuts, 1574 hours of raw Q-band data and 1208
hours of raw W-band data remained for scientific analysis. These
cuts rejected about 27$\%$ of Q-band data and 44$\%$ of W-band
data.

\subsection{Data Reduction}

In order to go from raw time ordered data to calibrated data, the
average offset per half scan, auto-calibration sequences, and time
domain spikes were removed, the data were calibrated, and binned
in azimuth.

Although the signal rms per azimuth bin was consistent with, but
slightly higher than, what the instrument sensitivity would predict,
calibrated data binned in azimuth did reveal the presence of a scan
synchronous signal (SSS). The SSS was sensitive to atmospheric
temperature changes and was likely due to thermal emission from the
dome into the beam sidelobes of the telescope. A linear regression
of calibrated signal as a function of azimuth, performed for each 20
minutes observing period, revealed an average slope of 1 mK/degree
in both bands, with variations up to 5 mK/degree/day and 14
mK/degree/day in Q and W bands respectively. Several strategies were
tested to minimize SSS contribution, but no reliable model could be
established to describe the SSS as a function of ambient
temperature, time, or housekeeping data. The SSS removal technique
employed involved subtracting from each file a second-order
polynomial fit to the calibrated data binned in azimuth. After this
procedure, pointing reconstruction was performed and the resulting
data was submitted to a maximum likelihood map making algorithm.

\subsection{Map Making}

We adopted the HEALPix\footnote{http://healpix.jpl.nasa.gov/}
pixelization scheme \citep{gorski05}, and used the
MADCAP\footnote{http://crd.lbl.gov/$\sim$borrill/cmb/madcap/}
package to produce Stokes Q maps \citep{borrill99}. The required
inverse time-time noise correlation matrix in the frequency domain
was computed as described in \citet{stompor02}. Noise maps were
estimated by computing the difference between two maps, each one
produced with half of the available data. Finally, after removing
any residual offset by imposing the condition that the map has zero
mean, a $\chi^{2}$ test for statistical significance was performed
on all the sky and noise maps to check if they were consistent with
the null hypothesis, i.e. consistent with no signal.

For the Q-band maps, the null hypothesis was rejected in all the sky
and noise maps produced when including data acquired during the day;
these maps contain non-stationary residuals from spurious signals
due to SSS and the Sun. We applied several masks to select
sub-regions in the maps that may be less affected by SSS but the
null hypothesis was also rejected in all these sub-regions. For maps
obtained using only data acquired during the night, the null
hypothesis was accepted in all the sky and noise maps produced. The
same result was also obtained when applying masks to select
sub-regions in these maps.

For the W-band maps, the situation was different. The W-band data
suffered from much higher SSS contamination, and we were unable to
determine a filter process that could retain sky signal and remove
the SSS. Because of this, these maps were rejected during further
analysis.

As a result, the Q-band night time data is the best data set
available from WMPol. The Stokes Q map obtained with this subset
contains 422 hours of integration, and covers an area of 3
$\mathrm{deg}^{2}$ in a region near the NCP with an effective
resolution of $24\arcmin$ FWHM at 42 GHz. The Stokes Q map contains
231 pixels (HEALPix Nside=512, which corresponds to a 6.9 arcmin
pixel size), an error of 45 $\mu\mathrm{K}/\mathrm{pixel}$, an rms
of 76 $\mu\mathrm{K}$, and thus an error in the mean of 5
$\mu\mathrm{K}$. These 422 hours of data were collected between 2004
July 6 and October 17 and comprise $17\%$ of the possible observing
time during this 104 day period. The day time data that were
collected during this time period and passed all cuts but were
ultimately not included in the final map comprise 378 additional
hours.

\subsection{Angular Power Spectra}

\begin{figure}
\plotone{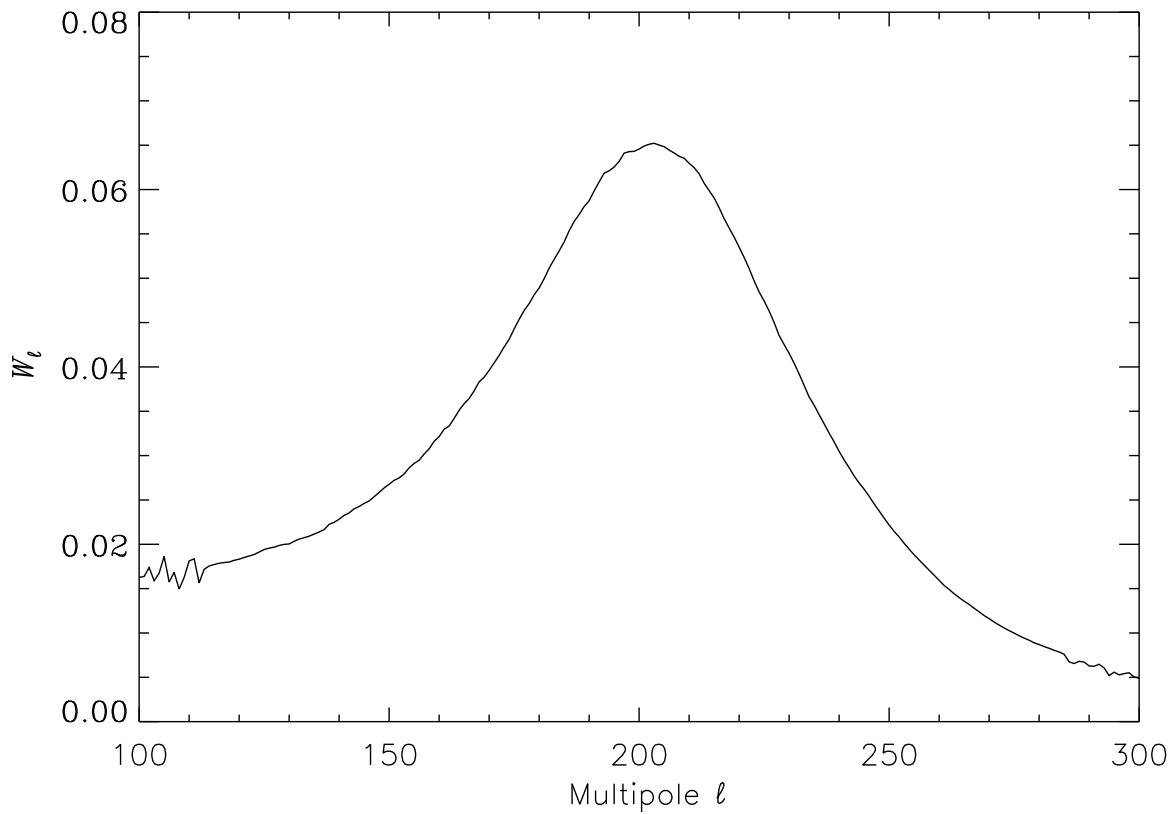} \caption{The WMPol transfer function for Q-band.
The transfer function is noisy for $\ell<120$ and it was not
computed for $\ell<100$. The multipole $\ell=120$ corresponds to
an angular scale of about $1\fdg5$, which is slightly larger than
the radius of the sky area observed. The transfer function becomes
negligible for $\ell>300$, which corresponds to an angular scale
of about $0\fdg6$, a value that approaches the Q-band beam size.
\label{windowfunction}}
\end{figure}

\begin{figure}
\plotone{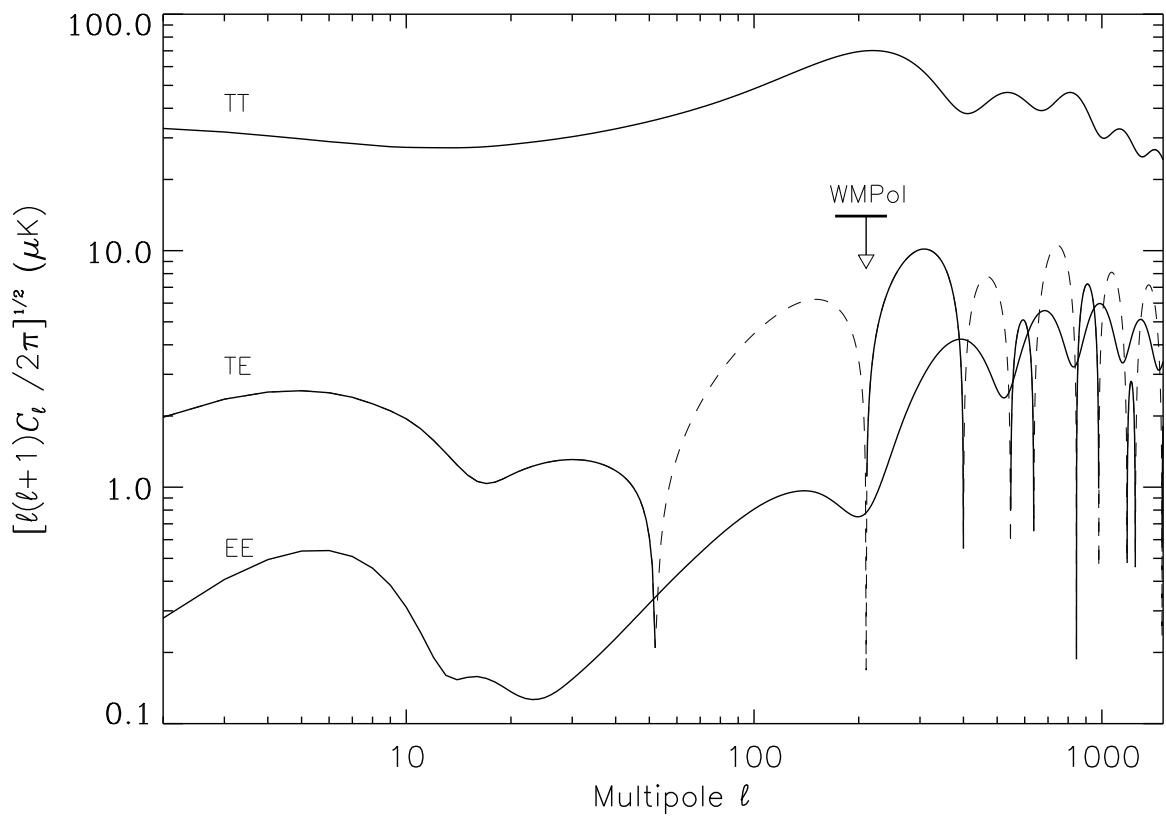} \caption{WMPol upper limit of 14 $\mu\mathrm{K}$
($95\%$ confidence limit), on $E$-mode polarization, in the
multipole range $170<\ell<240$. This result was obtained with 422
hours of observations of a 3 $\mathrm{deg}^2$ sky area near the
North Celestial Pole. The predicted angular power spectra $TT$,
$TE$, and $EE$ from a $\Lambda$-CDM concordance model are also
plotted for illustration. \label{upperlimit}}
\end{figure}

From Monte Carlo simulations, the WMPol \emph{EE} transfer
function was computed assuming a Gaussian beam, the scan strategy,
the pixelization scheme, the implemented data reduction pipeline,
and the data analysis package. The transfer function is the ratio
between the output pseudo angular power spectrum and the
corresponding input angular power spectrum. This procedure allowed
us to compute the transfer function in the multipole range
$100<\ell<500$. The WMPol transfer function is noise dominated for
$\ell<120$, it becomes negligible for $\ell>300$, and it takes
into account that we have observed only the $Q$ Stokes parameter.
Hence WMPol is sensitive to the multipole range
$170\lesssim\ell\lesssim240$. We estimated $EE$ using the PolSpice
package \citep{chon04,szapudi01}, which is a method for estimating
angular power spectrum from the two-point correlation function and
deals with the effects of limited sky, beam smoothing, noise
contamination, and inhomogeneous pixel errors. We obtained an
upper limit of 14 $\mu\mathrm{K}$ with 95$\%$ confidence level.
This upper limit does not include the uncertainties in the Q-band
calibration ($20\%$), beam size ($13\%$, which corresponds to
$26\%$ uncertainty in beam solid angle), or beam shape, which was
assumed to be Gaussian. The Table~\ref{t8} compares our result to
recent CMB polarization measurements, Figure~\ref{windowfunction}
displays the WMPol transfer function and Figure~\ref{upperlimit}
shows our result in the context of the $\Lambda$-CDM concordance
model.

\section{Conclusion}

WMPol was designed, built, and tested at University of California,
Santa Barbara and installed in the WMRS Barcroft Observatory to
measure CMB polarization in Q-band and W-band. With this telescope
and receiver system 2169 hours of data were collected during 2004.
The WMPol telescope proved to be robust and a novel remote control
system using relatively inexpensive and commercially-available
hardware and software was successfully implemented and utilized to
operate the telescope over the internet.

WMPol sets an upper limit on $E$-mode polarization of 14
$\mu\mathrm{K}$ ($95\%$ confidence limit, not including calibration
and beam uncertainties) in the multipole range $170<\ell<240$. This
result was obtained from 422 hours of observations of a Q-band
continuous comparison polarimeter on a 3 $\mathrm{deg}^2$ sky area
near the North Celestial Pole. The reported upper limit is based on
a method that estimates the CMB polarization power spectra via the
two-point correlation function, and it is consistent with $EE$
polarization predicted from a standard $\Lambda$-CDM concordance
model.

\acknowledgments

This work would not have been possible without the support of Frank
Powell, the Director of the White Mountain Research Station, and the
dedicated WMRS staff including Associate Director John Smiley, Paul
Addison, Dori Cann, Martin Freeman, Ryan Kitts, Matt Lokken, Richard
McDade, Gary Milender, DonnaLisa Shinn, Valerie Thorp, Mike Virgin,
and Denise Waterbury. We are also grateful for the many UCSB
undergraduate students who contributed during the course of
building, testing, and installing the telescope. The superb UCSB
Machine Shop was instrumental in the success of this project. We
thank Todd Gaier and Sandy Weinreb for providing the HEMT
amplifiers.  We also acknowledge the efforts of Chris O'Dell who did
much of the initial work in designing the W-band polarimeter and
Miikka Kangas who prepared some of the figures in this article. We
thank Julian Borrill, Radek Stompor, and Chris Cantalupo for help
with and providing resources to use MADCAP and we thank Eric Hivon
and Ian O'Dwyer for help with and providing resources to use
PolSpice.

This work was supported by NSF grants AST-9802851, 9813920, and
0118297 and the UCSB Office of Research. The development of the
carbon fiber reflectors was supported by NASA grants NAG5-4078,
NAG5-9073, and NAG5-4185. We are especially grateful to Thin Film
Technology, Inc., Buellton, CA for coating the reflectors. The
W-band feeds were developed at the the Microwave Laboratory of
Istituto di Fisica dei Plasmi - CNR, Milan, Italy.

A.L. was supported by a California Space Grant Consortium Graduate
Research Fellowship and A.L., N.S., and B.W. were supported by the
WMRS Graduate Student Research Fellowship. R.L. was supported by
CAPES and the NASA Planck project under JPL contract \#1261740.
T.V. and C. A. W. were partially supported by FAPESP grant
00/06770-2. In addition, T.V. acknowledges support from CNPq grant
305219/2004-9 and C. A. W. acknowledges support from CNPq grant
307433/2004-8-FA and FAPESP grant 96/06501-4. G. M. was supported
by a UCSB Chancellor's Undergraduate Research Award and M. vdH.
was supported by a UCSB Undergraduate Research and Creative
Activities Award.

Some of the results in this paper have been derived using the
HEALPix package \citep{gorski05}. The research described in this
paper was performed in part at the Jet Propulsion Laboratory,
California Institute of Technology, under a contract with the
National Aeronautics and Space Administration.

\clearpage

\begin{deluxetable}{lrr}
\tablewidth{0pt}
 \tablecaption{Design of the WMPol Corrugated Scalar Feed Horns \label{t1}} \tablehead{\colhead{Parameter} &
\colhead{Q-band} & \colhead{W-band}} \startdata

Flare angle (degrees) & 7 & 7 \\
Aperture diameter (mm) & 27.16 & 12.08 \\
Aperture corrugation depth (mm) & 1.9 & 0.88 \\
Aperture corrugation width (mm) & 1.45 & 0.67 \\
Throat diameter (mm) & 9.04 & 2.972 \\
Throat corrugation depth (mm) & 3.55 & 1.51 \\
Throat corrugation width (mm) & 1.45 & 0.13 \\
Number of corrugations & 34 & 37 \\
Corrugation step (mm) & 2.17 & 1.00 \\
Design return loss (dB) & $< -35$ & $< -35$ \\
FWHM (degrees) & 20 & 20 \\
Side lobe level(dB) & $< -30$ & $< -30$ \\

\enddata
\end{deluxetable}

\begin{deluxetable}{lrr}
\tablewidth{0pt}
 \tablecaption{Specifications of the Orthomode Transducers \label{t2}} \tablehead{\colhead{Parameter} &
\colhead{Q-band} & \colhead{W-band}} \startdata

Passband (GHz)& $36.0-43.0$\tablenotemark{*} & $85.0-105.0$ \\
VSWR & $1.2:1$ Max & $1.5:1$ Max \\
Insertion loss (dB) & $0.1$ Max & Not specified \\
Cross polarization (dB) & $-35$ & $-30$ \\

\enddata
\tablenotetext{*}{useful band of Q-band OMT was measured to extend
to 46 GHz}

\end{deluxetable}

\begin{deluxetable}{lll}
\tablewidth{0pt}
 \tablecaption{Major Elements of the Polarimeters \label{t3}} \tablehead{\colhead{Device} &
\colhead{Source (Q-band)} & \colhead{Source (W-band)}}
\startdata Corrugated scalar feed horns & Univ. of Milan (Italy) & INAF-IASF (Italy) \\
Orthomode transducers & Custom Microwave\tablenotemark{a} & Vertex RSI\tablenotemark{b} \\
Hybrid couplers & QuinStar Technology\tablenotemark{c} & Millitech\tablenotemark{d} \\
Cryogenic amplifiers & JPL/UCSB & JPL \\
Warm amplifiers & Avantek & JPL \\
Phase switches & Pacific Millimeter Products\tablenotemark{e} &
Pacific Millimeter Products \\
Adjustable attenuators & Aerowave\tablenotemark{f} & Aerowave\\
Bandpass filters & UCSB & UCSB \\
Microwave detector diodes & Pacific Millimeter Products & Pacific
Millimeter Products\\
Differential amplifier & UCSB & UCSB\\

\enddata
\tablenotetext{a}{Custom Microwave, Inc., Longmont, CO 80501,
http://www.custommicrowave.com} \tablenotetext{b}{Vertex RSI,
Torrance, California 90505, http://www.tripointglobal.com}
\tablenotetext{c}{QuinStar Technology, Inc., Torrance, California
90505, http://www.quinstar.com} \tablenotetext{d}{Millitech, Inc.,
Northampton, MA 01060, http://www.millitech.com}
\tablenotetext{e}{Pacific Millimeter Products, Golden, CO 80401,
http://www.pacificmillimeter.com} \tablenotetext{f}{Aerowave, Inc.,
Medford, MA, 02155, http://www.aerowave.net/}

\end{deluxetable}

\begin{deluxetable}{ccrc}
\tablewidth{0pt}
 \tablecaption{Parameters of the Polarimeters \label{t4}} \tablehead{ \colhead{} &
\colhead{$T_{system}$} & \colhead{$\Delta\nu_{eff}$} &
\colhead{Sensitivity}\\\colhead{Channel} & \colhead{(K)} &
\colhead{(GHz)} & \colhead{(mK$\cdot\mathrm{s}^{1/2}$)}}
 \startdata Q-band (38-46 GHz) & 127 & 8.0 & 3.5 \\ W-band (82-98 GHz) & 120 & 12.6 & 3.0 \\

\enddata
\end{deluxetable}

\begin{deluxetable}{lr}
\tablecaption{Primary Reasons for Not Taking Data \label{t5}}
\tablehead{\colhead{Issue} & \colhead{Hours}} \startdata \\Weather\tablenotemark{*} \space (snow, hail, rain, high wind) & 615 \\
Problems with remote control system (often caused by lightning) & 955 \\ Scheduled power outages & 71 \\
Cryocooler recycling & 147 \\ Various glitches (computer/code
crashes, cryocooler overheating, etc.) & 121 \\ Telescope
maintenance and manual calibrations & 194\\ \\ \tableline\\
Total lost observing time & 2103\\
Total successful observing time (Raw data) & 2169\\
Total possible observing time & 4272
\enddata \tablenotetext{*}{Only
includes hours where weather was the primary cause of not taking
data. There were $\sim 1300$ hours total during the observing
campaign when the dome shutter was required to be closed due to bad
weather.}
\end{deluxetable}

\begin{deluxetable}{lcccc}
\tablewidth{0pt} \tablecaption{WMPol Beam Model Characteristics
\label{t6}}

\tablehead{\colhead{Parameter}&  \colhead{Q-band}   &
\colhead{W-band} }

\startdata

DIR (dBi)    & 55.65 & 62.62    \\
XPD (dB)     & 39.81 & 67.01    \\
\% DEPOL     & 0.03 & 5.6E-05  \\
FWHM X (arcmin) & 19.08 & 8.45     \\
FWHM Y (arcmin) & 18.61 & 8.45   \\
FWHM AVE (arcmin)& 18.845 & 8.45    \\
ELL       & 1.03 & 1.0    \\

\enddata

\tablecomments{DIR is the maximum directivity of the co-polar
component. XPD is the cross polar discrimination defined as the
ratio between the copolar maximum and the cross polar maximum. \%
DEPOL is the integrated depolarization factor defined as
$(1-\frac{\sqrt{Q^2+U^2+V^2}}{I}) \times 100$ where Q, U, V, and I
are the Stokes parameters. FWHM X and FWHM Y are the angular
resolutions along the major and minor axis of the elliptical fit,
ELL is the ellipticity of the -3 dB contour level}

\end{deluxetable}

\begin{deluxetable}{lrr}
\tablecaption{Cuts of Raw Data \label{t7}}
\tablehead{\colhead{Issue} &
\colhead{Q-band (hours)} & \colhead{W-band (hours)}} \startdata Raw data & 2169 & 2169 \\\\ \tableline \\
Not scanning (testing, calibration, sources) & 130 & 130 \\
Short files (less than 60 s) & $<$ 1 & $<$ 1 \\
Saturation (due to Sun) & 188 & 468 \\
RMS cut & 173 & 295 \\
Cold plate temperature & 12 & 6\\
Weather image & 92 & 62\\\\ \tableline \\
Total cut & 595 & 961 \\\\ \tableline \\
Data remaining & 1574 & 1208 \\
\enddata
\end{deluxetable}

\begin{deluxetable}{ccccc}
\tablecaption{Comparison between recent CMB polarization
experiments \label{t8}}\tablehead{\colhead{Multipole range} &
\colhead{Frequency (GHz)} & \colhead{Main $EE$ result} &
\colhead{Experiment} & \colhead{Site}} \startdata
$66<\ell<505$ & $90$ & $<14\ \mu\mathrm{K}$ & PIQUE & NJ, USA \\
& & & \cite{hedman01} &  \\ \tableline
$2<\ell<20$ & $26-36$ & $<10\ \mu\mathrm{K}$ & POLAR & WI, USA \\
& & & \cite{keating01} &  \\ \tableline
$59<\ell<334$ & $90$ & $<8.4\ \mu\mathrm{K}$ & PIQUE & NJ, USA \\
& & & \cite{hedman02} &  \\ \tableline
$30\lesssim\ell\lesssim 900$ & $26-36$ & Detection & DASI & South Pole\\
& & & \cite{kovac02} &  \\ \tableline
$95<\ell<555$ & $26-36$ & $<32\ \mu\mathrm{K}$ & COMPASS & WI, USA  \\
& & & \cite{farese04} & \\ \tableline
$400\lesssim \ell\lesssim 1500$ & $26-36$& Detection & CBI & Chile \\
& & & \cite{readhead04} &  \\ \tableline
$640<\ell<1270$ & $90$ & Detection & CAPMAP & NJ, USA \\
& & & \cite{barkats05} &  \\ \tableline
$446<\ell<779$ & $26-36$ & $<7\ \mu\mathrm{K}$ & CBI & Chile \\
$930<\ell<1395$ &  & $<12.8\ \mu\mathrm{K}$ & \cite{cartwright05} & \\
$1539<\ell<2702$ &  & $<25.1\ \mu\mathrm{K}$ & &  \\
\tableline
$201<\ell<1000$ & $145$ & Detection & BOOMERANG & Antarctica \\
& & & \cite{montroy06} &  \\ \tableline
$2<\ell<6$ & $23-94$ & Detection & WMAP & L2 \\
& & & \cite{page07} &  \\ \tableline
$151<\ell<693$ & $140$ & Detection & MAXIPOL & NM, USA \\
& & & \cite{wu_jh07} &  \\ \tableline
$170<\ell<240$ & $42$ & $<14\ \mu\mathrm{K}$ & WMPol & CA, USA \\
& & & This work &  \\
\enddata
\end{deluxetable}

\clearpage

\clearpage

\end{document}